\documentclass[12pt]{article}
\usepackage{a4,graphicx}
\usepackage[utf8]{inputenc}
\usepackage{amsmath,amsthm,amsfonts,amssymb,amsopn,amscd}     

\usepackage{verbatim}
\usepackage{amssymb}
\usepackage{amsbsy}
\usepackage{amscd}
\usepackage{amsmath}
\usepackage{amsthm}
\usepackage[mathscr]{eucal}
\usepackage{graphicx}

\newcounter{allcount} \numberwithin{allcount}{section}

\newtheorem{lemma}[allcount]{Lemma}
\newtheorem{propo}[allcount]{Proposition}
\newtheorem{coro}[allcount]{Corollary}
\newtheorem{remark}[allcount]{Remark}

\def\inv{^{-1}}
\def\eps{\varepsilon}
\def\sig{\sigma}

\def\tr {{\mathrm{tr}}}
\def\id {{\mathrm{id}}}
\def\Ad {{\mathrm{Ad}}}
\def\uni {_{\mathrm{uni}}}
\def\Mat {{\mathrm{Mat}}}
\def\Rep {{\mathrm{Rep}}} \def\Conj {{\mathrm{Conj}}}

\def \RR {{\mathbb R}}
\def \QQ {{\mathbb Q}}
\def \CC {{\mathbb C}}
\def \ZZ {{\mathbb Z}}
\def \NN {{\mathbb N}}
\def \eins {{\mathbf 1}}
\def \End {{\mathrm{End}}}
\def \Hom {{\mathrm{Hom}}}
\def \Mor {{\mathrm{Mor}}}
\def\DHR{\mathrm{DHR}}
 \def\RA{\Rightarrow}
 \def\ol{\overline}

 \def\SU{\mathrm{SU}}

\def\bpm{\begin{pmatrix}} \def\epm{\end{pmatrix}}
\def\sumno{\sum\nolimits}
\def\qed{\hfill$\square$}
\newcommand{\uphook}{\rotatebox[origin=c]{25}{$\hookrightarrow$}\mkern2mu}
\newcommand{\downhook}{\rotatebox[origin=c]{-25}{$\hookrightarrow$}\mkern2mu}

\newcommand{\be}{\begin{equation}} 
\newcommand{\ee}{\end{equation}}
\newcommand{\ba}{\begin{array}} 
\newcommand{\ea}{\end{array}}
\newcommand{\bea}{\begin{eqnarray}} 
\newcommand{\eea}{\end{eqnarray}}
\newcommand{\eref}[1]{Eq.~(\ref{#1})}
\newcommand{\sref}[1]{Sect.~\ref{#1}}

\newcommand{\cref}[1]{Cor.~\ref{#1}}
\newcommand{\lref}[1]{Lemma~\ref{#1}}

\newcommand{\pref}[1]{Prop.~\ref{#1}}

\begin{document}


\title{{\bf The hypergroupoid of
  boundary conditions \\ for local quantum observables}\footnote{Talk
    of the second author at the 9th Seasonal Institute
  of the Mathematical Society of Japan
``Operator Algebras and Mathematical Physics'', Sendai, Japan, August
2016, submitted to the proceedings.}}


\author{Marcel Bischoff \\ \small Vanderbilt University,
  Nashville, TN, USA \\[-1mm] \small E-mail: {\tt
    marcel.bischoff@vanderbilt.edu} \\[5mm] Karl-Henning Rehren \\ \small
Institut f\"ur Theoretische Physik, Universit\"at G\"ottingen,
\\[-1mm] \small Friedrich-Hund-Platz 1, D-37077 G\"ottingen, Germany
\\[-1mm] \small E-mail: {\tt rehren@theorie.physik.uni-goettingen.de}}




\maketitle


\begin{abstract}
We review the definition of hypergroups by Sunder, and we associate a
hypergroup to a type III subfactor $N\subset M$ of finite index, whose
canonical endomorphism $\gamma\in\End(M)$ is multiplicity-free. It is
realized by positive maps of $M$ that have $N$ as fixed points. If the
depth is $>2$, this hypergroup is different from the hypergroup
associated with the fusion algebra of $M$-$M$ bimodules that was
Sunder's original motivation to introduce hypergroups.  

We explain how the present hypergroup, associated with a suitable
subfactor, controls the composition of transparent boundary conditions
between two isomorphic quantum field theories, and that this
generalizes to a hypergroupoid of boundary conditions between
different quantum field theories sharing a common subtheory.  
\end{abstract}

\section{Introduction}
\label{s:intro}

\subsection{Boundary conditions in quantum field theory}
\label{s:i-qft}

A boundary condition is the specification of algebraic relations
between the local and covariant quantum observables (``fields'') at
the two sides ($L=$ left, $R=$ right) of a boundary in Minkowski
spacetime (e.g., the surface $x=0$). The quantum fields on both sides
are assumed to share the fields of a common subtheory $A$ defined in
the entire spacetime, so that the boundary is ``transparent'' for $A$. 
Heuristically, the quantum field theory $A$ is assumed to contain the
stress-energy tensor whose moments are the generators of the
Poincar\'e group. This property ensures (and in chiral and
two-dimensional conformal QFT is equivalent) that energy and momentum
are conserved at the boundary. It also implies that the theories $B^L$
and $B^R$, defined only on their halfspaces, actually extend to the
entire spacetime. A boundary condition may therefore be thought of as
an algebraic relation between the fields defined on one side of the
boundary and extended to the other side, and the field originally
defined on the other side.   

A boundary condition may be just a gauge transformation, when $B^L$
and $B^R$ are isomorphic and admit a global gauge symmetry such that
the elements of $A$ are gauge invariant. However, the general case
exhibits a much richer structure due to the existence, in low spacetime
dimensions, of superselection sectors (representations of $A$) that
are not related to some gauge symmetry. 

An analysis of possible transparent boundary conditions under the
stated assumptions has been initiated \cite{BKLR}, leading to a
``universal construction'' of a reducible boundary condition whenever
$A\subset B^L$ and $A\subset B^R$ are given, along with a
classification of irreducible boundary conditions. It was found that
boundary conditions are in general quadratic relations, that reduce to
linear ones only in special cases, including the case of gauge
transformations.  

In this note, we want to show that boundary conditions correspond to
the elements of a hypergroup, and its composition law describes a 
``composition of boundary conditions''. When the observables $B$ in
each region of spacetime separated by several boundaries are not
isomorphic, the hypergroup is to be replaced by a hypergroupoid. Consider the 
juxtaposition of two boundaries, say, at $x=-a$ and $x=+a$ with
observables $B^L$ at $x<-a$, $B^M$ at $-a<x<a$, and $B^R$ at $x>a$. 
The composition of boundary condition computes the
possible boundary conditions between $B^L$ and $B^R$, that are
compatible with two given boundary conditions between $B^L$ and $B^M$
and between $B^M$ and $B^R$. 

It shall become clear from our analysis that the hypergroup(oid)s describing the
possible boundary conditions depend only on the representation
category $\DHR(A)$ of the underlying common subtheory and its
Frobenius algebras (Q-systems) determining the extensions. A few examples
are listed in the end of \sref{s:jux}.

\subsection{Hypergroups}
\label{s:i-hg}

According to the definition given by Sunder and Wildberger in \cite{SW}, a
(finite) hypergroup is a (finite-dimensional) vector space with a
distinguished basis $k_i$ satisfying an associative multiplication
\bea\label{hg}
k_ik_j = \sumno_{\ell} \lambda_{ij}^\ell\cdot k_\ell
\eea
with non-negative real coefficients $\lambda_{ij}^\ell$ such that, for each $i,j$
\bea\sumno_\ell \lambda_{ij}^\ell = 1.
\eea
There is a neutral element $k_0=1$ and an involution
$i\leftrightarrow \ol i$ such that $\lambda_{ij}^0\neq0$ if and only if
$j=\ol i$. In the discrete infinite case, for each $i,j$ only finitely many
$\lambda_{ij}^\ell$ are allowed to be nonzero. 

\newpage

This definition clearly generalizes the notion of a group, by admitting
the product of two basis elements to be a convex sum of basis
elements. Fusion algebras 
\bea\label{fa}
f_if_j = \sumno_{\ell} n_{ij}^\ell\cdot f_\ell
\eea
with $n_{ij}^\ell\in\NN_0$, equipped with a conjugation such that
$n_{ij}^0 = \delta_{j\ol i}$, also define hypergroups \cite{S} by putting
$k_i:=\dim_i\inv\cdot f_i$, where $\dim_i\geq 1$ is the Perron-Frobenius dimension
of $f_i$. In particular, also the duals of compact groups are
hypergroups. 

Fusion algebras were Sunder's original definition of
hypergroups in \cite{S}, where he had in mind the fusion algebra of
$M$-$M$ bimodules of a (finite-index finite-depth type II$_1$)
subfactor $N\subset M$. There, he also showed the uniqueness of the
dimension function $f_i\to \dim_i\in\RR_+$. As a consequence of the
integrality of $n_{ij}^\ell$, the coefficients $\lambda_{ij}^\ell$ of
a finite hypergroup must be algebraic numbers. The latter condition
was, however, dropped in the definition in \cite{SW}. This
is a proper generalization, since every $\lambda\in(0,1)$ defines a two-element
hypergroup by $k_1^2=\lambda\cdot k_0+(1-\lambda)\cdot k_1$, whereas
according to the definition in \cite{S}, $\lambda$ must be of the form
$x=1+\frac12r-r\sqrt{\frac14+r\inv}$ with $r=(n_{11}^1)^2/n_{11}^0\in\QQ$.

Examples of the more general definition in \cite{SW} include also the
hypergroups $\Conj(G)$ of the (properly normalized sums over the
elements of) conjugacy classes of a finite group, or double cosets of
a finite group with respect to some subgroup. These arise by rescaling
relations of the form \eref{fa} where, however, possibly $n_{i\ol i}^0>1$.   

The weight $\mu_i\geq1$ of an element $k_i$ of a hypergroup is by definition the
inverse of the coefficient $\lambda_{i\ol i}^0$. In particular, for the
hypergroup associated with a fusion algebra, 
\bea\label{fwei}
\mu_i=(\dim_i)^2.
\eea

The Haar measure of a hypergroup is the convex combination
\bea\label{haar}
H=\big(\sumno_i \mu_i\big)\inv\sumno_i \mu_i k_i.
\eea
It satisfies $H=\ol H = H^2 =Hk_i=k_iH$ for all $i$. 

In the present note, we adhere to the latter, more general definition
of a hypergroup as in \cite{SW}, and associate another natural hypergroup
$K_{N\subset M}$ to a finite-index subfactor \cite{B}, different from
the one coming from the fusion algebra of $M$-$M$ bimodules when the
depth is $>2$.  

Our definition requires a condition of multiplicity-freeness: If
$M\subset M_1$ is the Jones basic construction of $N\subset M$, 
then $M_1$ as an $M$-$M$ bimodule is assumed to decompose into
irreducible bimodules without multiplicities. This is the case for
subfactors $N=M^G\subset M$ given by the fixed points under an outer
action of a group $G$ (where the $M$-$M$ bimodules are in one-to-one
correspondence with the elements $g\in G$), but also for many other subfactors.

\subsection{The hypergroupoid of boundary conditions}
\label{s:i-bc}

Our motivation comes from quantum field theory. We are 
interested in the local subfactors $A(O)\subset B(O)$, where $A(O)$
and $B(O)$ are the algebras of observables localized in a spacetime
region $O$ of a QFT $B$ and a subtheory $A$. For suitable regions
(double cones or intervals), the structure of this subfactor does not
depend on $O$, and it completely determines the extension $A\subset B$
of the quantum field theories. (Because we want to concentrate on the
subfactor aspects, we do not record in detail the necessary technical 
assumptions \cite{BKLR,LR}, that are satisfied by large classes of
(conformal) quantum field theories.) Because the structure is independent of
the region, $O$ will be indicated below only for the sake of definiteness. 

Since the local algebra are of type III, the Jones theory of subfactors
has to be appropriately adapted \cite{KL,L}. Instead of $X$-$Y$
bimodules (where $X$ and $Y$ may be either $N$ or $M$), one deals with
homomorphisms $Y\to X$. In particular, in the quantum field theory
context, the $N$-$N$ bimodules correspond to DHR endomorphisms
of $A$ localized in $O$, considered as endomorphisms of $N=A(O)$. The
irreducible endomorphisms $\gamma_a$ contained in the canonical
endomorphism $\gamma$ of $M=B(O)$ replace the $M$-$M$ bimodules
contained in $M_1$, and may be thought of as generalizations of gauge
automorphisms.   

The subfactors $A(O)\subset B(O)$ fulfil the above-mentioned
multiplicity-freeness as a consequence of locality. The elements
$k_a$ of the associated hypergroup $K_{A\subset B}$ according to \cite{B}
are then realized by completely positive maps $\phi_a:B\to B$, whose
fixed points in $B(O)$ are the elements of $A(O)$. In special cases,
these positive maps may be automorphisms and coincide with
$\gamma_a$. Thus, the hypergroup is an appropriate generalized
symmetry concept for local extensions $A\subset B$.

We shall show in \sref{s:homo} that the elements of the hypergroup
$K_{A\subset B}$ characterize the algebraically admitted boundary
conditions between two isomorphic copies $B^L$ and $B^R$ of an
extension $B$ of $A$, regarded as quantum field theories defined to
the left and right of a boundary in space. We show in \sref{s:jux}
that the hypergroup composition law controls the composition of
boundary conditions, and in \sref{s:hetero}, we sketch the (immediate)
generalization to the general case when $B^L$ and $B^R$ may range over
several inequivalent local extensions of a (given) quantum field theory $A$.

\section{Subfactors}
\label{s:subf}
\setcounter{equation}{0}

We deal with type III subfactors, which is the appropriate setting
for the intended applications in quantum field theory. However, the
content of this and the next section can be literally translated into
the more popular type II$_1$ setting, by substituting ``bimodules''
for ``homomorphisms''. 

For two (type III) factors $A$, $B$ and a pair of homomorphisms
$\varphi_1,\varphi_2:A\to B$, intertwiners are the elements of the
linear spaces in $B$
$$\Hom(\varphi_1,\varphi_2)=\{t\in
B:t\varphi_1(a)=\varphi_2(a)t\;\forall a\in A\}.$$
Clearly, $t^*\in\Hom(\varphi_2,\varphi_1)$ iff
$t\in\Hom(\varphi_1,\varphi_2)$.  
The concatenation product of $t_1\in\Hom(\varphi_1,\varphi_2)$ and
$t_2\in\Hom(\varphi_2,\varphi_3)$ is the operator multiplication
$$t_2\circ t_1:= t_2\cdot
t_1\in \Hom(\varphi_1,\varphi_3).$$ 
For $\varphi_1,\varphi_3:A\to B$ and $\varphi_2,\varphi_4:B\to C$, the monoidal product of $t_1\in\Hom(\varphi_1,\varphi_3)$ and
$t_2\in\Hom(\varphi_2,\varphi_4)$ is 
\bea\label{monoi}
t_2\times t_1 := t_2\varphi_2(t_1)\equiv \varphi_4(t_1)t_2
\in\Hom(\varphi_2\varphi_1,\varphi_4\varphi_3).
\eea
With the monoidal product of homomorphisms
$\varphi_1\times\varphi_2 := \varphi_1\varphi_2$, one has a
C* two-category whose objects are factors, the one-morphisms are
homomorphisms, and the two-morphisms are intertwiners.  

For $\varphi_1,\varphi:A\to B$, we write $\varphi_1\prec\varphi$ if
there is an isometry $t\in\Hom(\varphi_1,\varphi)$. In the
type III case, every projection in $B$ can be written as $e=tt^*$ with an
isometry $t\in B$, and $\varphi_1(\cdot):=t^*\varphi(\cdot)t$ defines
$\varphi_1\prec\varphi$, regarded as the ``range'' of the projection
$e\in\Hom(\varphi,\varphi)$. Conversely, one may define the ``direct
sum'' of homomorphisms $\varphi_i:A\to B$ (unique up to unitary
equivalence) by choosing a partition of unity $1_B=\sum_ie_i$, writing
$e_i=t_it_i^*$, and defining $\varphi(\cdot)=\sum_it_i\varphi_i(\cdot)t_i^*$.   

\subsection{Conjugate endomorphisms}
\label{s:conj}

Two homomorphisms $\varphi:A\to B$ and $\ol\varphi:B\to A$ are 
conjugate to each other if there is a pair of intertwiners
$r\in\Hom(\id_A,\ol\varphi\varphi)$ and $\ol
r\in\Hom(\id_B,\varphi\ol\varphi)$ 
satisfying the conjugacy relations 
\bea
\label{conj} \qquad
(\eins_\varphi\times r^*)\circ (\ol r\times\eins_\varphi)=1_B,
\quad(\eins_{\ol\varphi}\times \ol r^*)\circ
(r\times\eins_{\ol\varphi})=1_A.
\eea
$r$ and $\ol r$ are multiples of isometries (because
$\Hom(\id,\id)=\CC\cdot 1$ for a factor) and can be normalized such
that 
\bea
\label{stand}
r^*\circ r=d\cdot \eins_A,\quad \ol r^*\circ\ol r=d\cdot \eins_B.
\eea
If $\varphi$ is irreducible, $d\geq1$ is unique, and is called the
dimension of $\varphi$. If $\varphi$ is reducible, $\dim_\varphi$ is
defined as the infimum of $d$ over all solutions to the conjugacy
relations \eref{conj}, and a solution $(r,\ol r)$ saturating the
infimum is called standard. One has
$(\dim_\varphi)^2=(\dim_{\ol\varphi})^2=[B:\varphi(A)]$, and 
the dimension is additive under direct sums and multiplicative under
composition of homomorphisms. 

Let $N\subset M$ be a type III subfactor of finite index
$[M:N]=d^2$. Let $\iota:N\hookrightarrow M$ be the embedding
homomorphism $\iota(n)=n\in M$, and 
$\ol\iota:M\to N$ a conjugate homomorphism, and $(w,v)$ a standard
solution to the conjugacy relations \eref{conj}. We call $\gamma:=\iota\ol\iota$
the canonical endomorphisms (of $M$), and $\theta:=\ol\iota\iota$ the
dual canonical endomorphisms (of $N$). Thus $w\in\Hom(\id_N,\theta)$
and $v\in\Hom(\id_M,\gamma)$, and $\dim_\iota=d$. Because $1_M=\iota(w^*)v$,
one has $m=1_Mm=\iota(w^*\ol\iota(m))v$ and hence
\bea
\label{gen}
M=\iota(N)\cdot v,
\eea
i.e., $v$ together with $N$ generates $M$. We shall refer to $v$ as
the canonical generator of $N\subset M$.  

The conjugation data define a conditional expectation $\mu:M\to N$
(unique in the irreducible case) 
\bea
\label{condex}
\mu(\cdot) = \dim_\iota\inv \cdot w^*\ol\iota(\cdot)w.
\eea

If $A=B^G$ are the fixed points under the outer action of a finite
group $G$, then $[B:A]=\vert G\vert$, $\dim_\iota=\vert
G\vert^{\frac12}$, $\Hom(\theta,\theta)$ is isomorphic to $\CC G$,
$\Hom(\gamma,\gamma)$ is commutative and isomorphic to $C(G)$. More
precisely, the irreducible sub-homomorphisms of $\gamma$ are the group
automorphisms $\alpha_g$, $g\in G$, and the condition expectation is
the group average
$$\iota\circ\mu(\cdot)=\vert G\vert\inv\sumno_{g\in G}\alpha_g(\cdot).$$

Frobenius reciprocity defines a bijective linear ``Fourier
transformation'' $\Hom(\theta,\theta)\to\Hom(\gamma,\gamma)$ by 
\bea
\label{fou}
\chi(t):= (v^*\times 1_\gamma)\circ(1_\iota\times t\times
1_{\ol\iota})\circ (1_\gamma\times v) = \chi(t^*)^* \hskip-10mm
\eea
with inverse 
\bea
\chi\inv(s) = (1_\theta\times w^*)\circ(1_{\ol\iota}\times s\times
1_{\iota})\circ (w\times 1_\theta) = \chi\inv(s^*)^*. \hskip-10mm
\eea
$\chi$ and $\chi\inv$ turn the concatenation products of the algebras
$\Hom(\theta,\theta)$ and $\Hom(\gamma,\gamma)$ into the ``convolution
products'' 
\bea
\label{conv}
t_1\ast t_2 &:=& \chi\inv(\chi(t_1)\circ\chi(t_2)), \\\notag  s_1\ast s_2 &:=&
\chi(\chi\inv(s_1)\circ\chi\inv(s_2)) 
\eea
with unit $ww^*$ respectively $vv^*$. 

(The terminology is justified because in the fixed point case $A=B^G$
with an abelian group $G$, $\chi$ is the usual Fourier transform between $G$ and
$\widehat G$ and $\ast$ is the usual convolution product.)

\subsection{Q-systems}

Given a subfactor $N\subset M$ and $w\in\Hom(\id_N,\theta)$ and
$v\in\Hom(\id_M,\gamma)$ as above, define
$x:=\ol\iota(v)\in\Hom(\theta,\theta^2)$. The data 
$$Q=(\theta,w,x)$$
with normalizations $w^*\circ w=(\dim_\theta)^{1/2}$, $x^*\circ
x=(\dim_\theta)^{1/2}\cdot 1_\theta$ satisfy the relations 
$$(1_\theta\times w^*)\circ x= 1_\theta = (w^*\times 1_\theta)\circ x,$$
$$(1_\theta\times x^*)\circ (x\times 1_\theta)= x\circ x^* =
(x^*\times 1_\theta)\circ (1_\theta\times x),$$
$$(1_\theta\times x)\circ x = (x\times 1_\theta)\circ x,$$
that define a Frobenius algebra or Q-system in the tensor category
$\End_0(N)$ (= the C* tensor category of endomorphisms of $N$ with
finite dimension). The Q-system encodes $N\subset M$ as an
extension of $N$ uniquely up to isomorphism \cite{L}. 

We shall freely use the relations of the Q-system, as well as the
conjugacy relations \eref{conj} throughout. 

In terms of the Q-system, the convolution product of
$\Hom(\theta,\theta)$ reads
\bea\label{ast}
t_1\ast t_2 = x^*\circ (t_1\times t_2)\circ x.
\eea

If $\theta$ belongs to a braided full subcategory of $\End_0(N)$ with
braiding $\eps$, then $Q$ is called commutative iff
$\eps_{\theta,\theta}\circ x=x$.  
\begin{propo}\label{p:mfree} If $\theta$ belongs to a braided
  subcategory of $\End_0(N)$ and $Q$ is commutative, then
$[\Hom(\theta,\theta),\ast]$ is commutative. Equivalently,
$[\Hom(\gamma,\gamma),\circ]$ is commutative. As a consequence, 
$\gamma$ is multiplicity-free:
$$\gamma\simeq\bigoplus\nolimits_a\gamma_a$$
with $\gamma_a$ irreducible and pairwise inequivalent. 
\end{propo}
{\em Proof:} The commutativity of the convolution product of
$\Hom(\theta,\theta)$ follows by
$$
x^*\circ (t_1\times t_2)\circ x = x^*\circ
\eps_{\theta,\theta}^*\circ(t_2\times t_1)\circ
\eps_{\theta,\theta}\circ x = x^*\circ (t_2\times t_1)\circ x
$$
where the second equality is the commutativity of $Q$. Hence
$\Hom(\gamma,\gamma)$ is commutative with respect to the ordinary =
concatenation product. Because the irreducible decomposition of
$\gamma$ is controlled by the projections in $\Hom(\gamma,\gamma)$,
the absence of multiplicities follows. \qed

\medskip

In the appendix, we record an unexpected quantization result for the
dimensions $\dim_{\gamma_a}$, that arises as a corollary of this fact.

\section{The hypergroup of a subfactor}
\label{s:hyperg}
\setcounter{equation}{0}

Let $N\subset M$ be an irreducible type III subfactor, and
$\theta=\ol\iota\iota$, $\gamma=\iota\ol\iota$ and
$w\in\Hom(\id_N,\theta)$, $v\in\Hom(\id_M,\gamma)$, and $Q=(\theta,w,x)$ as in \sref{s:subf}.

We assume that $\gamma$ is multiplicity-free. By \pref{p:mfree}, this
is certainly the case whenever the Q-system is commutative; but the 
existence of a braiding for $\theta$ is neither needed nor assumed in
this section.  

\subsection{Positive maps}
\label{s:pmaps}

Let $e_a\in\Hom(\gamma,\gamma)$ be the projections onto
$\gamma_a\prec\gamma$ and choose isometries
$s_a\in\Hom(\gamma_a,\gamma)$, such that $e_a=s_as_a^*$ and
$\sum_a s_as_a^*=1_M$. 

Then we decompose the conditional expectation \eref{condex},
considered as a positive map of $M$ into itself, as
$$
\iota\circ\mu(\cdot) = \frac1{\dim_\iota}  \cdot
\iota(w^*)\gamma(\cdot)\iota(w)
= \frac1{\dim_\iota}  \cdot \iota(w^*)\sumno_a s_a\gamma_a(\cdot)s_a^*\iota(w),
$$
where $s_a^*\iota(w)\in \Hom(\iota,\gamma_a\iota)$ are multiples of
isometries, $\iota(w^*)s_as_a^*\iota(w) =
\dim_{\gamma_a}/\dim_\iota$. In terms of isometries 
$h_a=(\dim_\iota/\dim_{\gamma_a})^{\frac12}s_a^*\iota(w)$, 
\bea\label{deco}
\iota\circ\mu(\cdot) = 
\frac1{\dim_\gamma} \sumno_a \dim_{\gamma_a}\cdot h_a^*\gamma_a(\cdot)h_a =:
\frac1{\dim_\gamma} \sumno_a \dim_{\gamma_a} \cdot \phi_a(\cdot). \hskip-17mm
\eea
\begin{propo}\label{p:hg} \cite{B} (i) The unital positive maps
\bea
\phi_a(\cdot)=h_a^*\gamma_a(\cdot)h_a:M\to M
\eea
are $N$-linear maps (i.e., $\phi_a(n_1mn_2)=n_1\phi_a(m)n_2$) with
fixed points $N$. \\ (ii) $\phi_a$ form a hypergroup  
$K_{N\subset M}$ under composition, with coefficients given in
\eref{coeff}. \\ (iii) The Haar measure of $K_{N\subset M}$ is
$\iota\circ\mu$ as in \eref{deco}. In particular, the weights are 
\bea\label{hwei}
\mu_a=\dim_{\gamma_a}.
\eea
\end{propo}
{\em Proof:} (i) The $N$-linearity is immediate by the definition of
$\phi_a$ and implies the trivial action on $N$ by choosing $m=1$. 
Conversely, if $\phi_a(m)=m$ for all $a$, then 
$\iota\mu(m)=m$ by \eref{deco}, hence $m\in N$. \\ (ii) We have  
$$\phi_a\circ\phi_b(\cdot) = (\dim_\iota)^2\cdot \iota(w^*) s_{ab}^* \gamma_a\gamma_b(\cdot)s_{ab}\iota(w)
$$
where $s_{ab}\equiv (s_a\times s_b)\circ (1_\iota\times w\times 1_{\ol\iota}) 
\in\Hom(\iota\ol\iota,\gamma_a\gamma_b)$. Splitting $\gamma_a\gamma_b$
into irreducibles, 
$$\phi_a\circ\phi_b(\cdot) = (\dim_\iota)^2\cdot
\sumno_{\gamma_i\prec\gamma_a\gamma_b}\iota(w^*)s_{ab}^*t_i\gamma_i(\cdot)t_i^*
s_{ab}\iota(w) $$
where $t_i\in\Hom(\gamma_i,\gamma_a\gamma_b)$. The intertwiner
$s_{ab}^*\circ t_i$ is in
$\Hom(\gamma_i,\iota\ol\iota)=\Hom(\gamma_i,\gamma)$; hence it
vanishes if $\gamma_i$ is not contained in $\gamma$ (which is possible
if the depth is $>2$); and it is a multiple $\alpha_{ab}^c\cdot s_c$
of $s_c$ if $\gamma_i=\gamma_c\prec\gamma$. Thus the sum runs only
over the subsectors $\gamma_c$ of $\gamma$, and each summand is a
positive multiple of $\phi_c(\cdot)$:
\bea\label{coeff}
\phi_a\circ\phi_b=\sumno_c \lambda_{ab}^c\cdot
\phi_c\quad\hbox{with}\quad
\lambda_{ab}^c=\dim_\iota\cdot\vert\alpha_{ab}^c\vert^2.
\eea
Because all $\phi_c$ are unital maps, the convex property of the
coefficients is automatic. Moreover, because also
$\gamma_c\prec\gamma_a\gamma_b$, it follows that $\lambda_{ab}^0>0$
iff $b=\ol a$.   
\\ (iii) From the fact that $\phi_a(\iota(\cdot)) = \iota(\cdot)$, we
conclude $\phi_a\circ \iota\mu=\iota\mu$, and because the Haar measure
is the unique convex sum satisfying $\phi_a\circ H=H$ for all $a$, it
follows that $H=\iota\mu$. \qed   

If $\omega=\omega\circ\mu$ is an invariant state on $M$, then it follows from
(iii) that $\omega=\omega\circ\phi_a$ for all $a$. Thus, the positive
maps $\phi_a$ are stochastic maps with respect to $\omega$.
\begin{remark}
Compare the hypergroup $K_{N\subset M}$ with the hypergroup
associated with the fusion algebra of the $M$-$M$ bimodules of a
subfactor of finite depth as in \sref{s:i-hg}. In the type III 
case, the $M$-$M$ bimodules are the irreducible subsectors $\gamma_i$ of
arbitrary powers of $\gamma$, of which $\gamma_a\prec\gamma$ are only
a subset. $\gamma_a$ exhaust all $\gamma_i$ if the depth is $=2$. In
this case, the two hypergroups are the same.  
Indeed, the conflicting formulas \eref{fwei} and \eref{hwei}
coalesce in this case, because if a subfactor $N\subset M$ has depth 2
and $\gamma$ is multiplicity-free (e.g., in the case of a fixed-point
subfactor), then all $\gamma_a\prec\gamma$ have dimension 1 ($\gamma_a$ are
automorphisms of $M$). \end{remark}
{\em Example:} The Goodman-de~la~Harpe-Jones subfactor of index $3+\sqrt3$
\cite{GHJ} has $\gamma=\id_M+\gamma_1$ where $\gamma_1$ has dimension
$d_1=2+\sqrt3$. Thus, the associated hypergroup 
$K_{\rm GHJ}$ is given by $\phi_1^2=d_1\inv(\phi_0+(d_1-1)\phi_1)$. 
This is not the hypergroup of any fusion algebra $f_1^2=f_0+nf_1$ with
$n\in\NN$, cf.\ \cite[Sect.~4.6]{B}. But it happens to be a quotient
of the hypergroup of the fusion algebra of the $M$-$M$ bimodules (the
irreducible subsectors of powers of $\gamma$) that was computed
in \cite[Table 3]{K}. This is always true for hypergroups
associated with commutative Q-systems in modular tensor categories
\cite[Thm.~ 5.16]{B}, but not in general: e.g., the dual of the GHJ
subfactor has the same hypergroup $K_{M\subset M_1}\cong K_{N_1\subset
  N}\cong K_{\rm GHJ}$, but it is not a quotient of the hypergroup of
the fusion algebra of the $N$-$N$ bimodules (the 
subsectors of powers of $\theta=\id_N+\theta_1$, \cite[Table 2]{K}). \\
More examples can be found at the end of \sref{s:jux}.   

\subsection{Matrix representations}
\label{s:reps}

A unitary matrix representation of a hypergroup $K$ is a
homomorphism $U$ of $K$ into $\Mat_m(\CC)$ such that $U(\ol
k)=U(k)^*$. (This does {\em not} mean that $U(k)$ are unitary
matrices.) We obtain the irreducible matrix representations of the
hypergroup $K_{N\subset M}$ as follows.  

Let $\theta_n\prec\theta$ be irreducible, 
arising with multiplicities $m_n=\dim\Hom$ $(\theta_n,\theta)$. By
Frobenius reciprocity
$\dim\Hom(\theta_n,\theta)=\dim\Hom(\iota,\iota\theta_n)$. Hence there
are bases of $m_n$ orthonormal isometries
$\psi_{n,i}\in\Hom(\iota,\iota\theta_n)$ ($i=1,\dots m_n$). They can
be written as $\psi_{n,i}=\iota(w_{n,i}^*)v$ with $w_{n,i}\in
\Hom(\theta_n,\theta)$, and
$w_{n,i}^*w_{n,j}=(\dim_\iota/\dim_{\theta_n})\cdot\delta_{ij}$
\cite{Tcats}. With $\theta_0=\id_N\prec\theta$ ($m_0=1$) and $w_0=w$, one has
$\psi_0=1_M$. 
\begin{propo}\label{p:rep}
The positive maps $\phi_a$ of \pref{p:hg} act linearly on the spaces
$\Hom(\iota,\iota\theta_n)$. The coefficients in 
\bea\label{rep}
\phi_a(\psi_{n,i})=\sumno_j \psi_{n,j}\cdot U_n(a)_{ji}
\eea
form irreducible unitary matrix representations $U_n$ of
$K_{N\subset M}$. 
\end{propo}
{\em Proof:} Obviously,
$h_a^*\gamma_a(\psi_{n,i})h_a\in\Hom(\iota,\iota\theta_n)$, and the
homomorphism property is automatic. The *-property follows from
$(\psi_{n,i})^* = \iota(r_n^*)\psi_{\ol n,i}$ with $r_n\in\Hom(\id_N,\ol
\theta_n\theta_n)$. \qed
\begin{propo}\label{p:pw}
The representations $U_n$ are irreducible and pairwise inequivalent. 
They exhaust the irreducible representations of $K_{N\subset M}$. 
\end{propo}
To prove \pref{p:pw}, we first state two Lemmas.
\begin{lemma}\label{l:rep}
The matrix coefficients $U_n(a)_{ji}=\psi_{n,j}^*\phi_a(\psi_{n,i})$
are given by  
$$U_n(a)_{ji}\cdot 1_\iota = \frac{\dim_{\iota}}{\dim_{\gamma_a}}\cdot (v^*\times 1_\iota)\circ(\iota\times
w_{n,j}w_{n,i}^*)\circ (e_a\times 1_\iota)\circ (1_\iota\times
w),$$
or equivalently 
\bea\label{U}\qquad 
U_n(a)_{ji}\cdot 1_{\theta_n} = (v^*\times s_a^*)\circ(1_\iota\times
w_{n,j}w_{n,i}^*\times 1_{\ol\iota})\circ (s_a\times v). 
\eea
\end{lemma}
{\em Proof:} The first equality follows by inserting
$\psi_{n,i}=\iota(w_{n,i}^*)v$ into the definition of
$\phi_a$. The equivalence is established by applying to the
expressions on the right-hand sides the traces 
$\tr_\iota(t)=v^*\circ t\circ v$ ($t\in\Hom(\iota,\iota)$) and
$\tr_{\theta_a}(t')=\ol r_a^*\circ t'\circ\ol r_a$
($t'\in\Hom(\gamma_a,\gamma_a)$) where $(r_a,\ol r_a)$ is a standard
solution to the conjugacy relations for $\gamma_a$ and
$\ol\gamma_a$, and exploiting the trace property \cite{LRo}. \qed

\medskip 

Notice that up to normalization, $U_n(a)_{ji}$ coincide with the
pairing between the projections $e_a\in\Hom(\gamma,\gamma)$ and the
matrix units of $\Hom(\theta,\theta)$ in \cite{WH}. There
are as many matrix units as there are projections $e_a$, namely
$\dim\Hom(\theta,\theta)=\dim\Hom(\gamma,\gamma) = 
\vert K_{N\subset M}\vert$. By similar standard computations using the
trace property \cite{LRo} as before, one finds
\begin{lemma}\label{l:uni} The square matrices
  $S^a_{n,i,j}=\frac{\sqrt{\dim_{\gamma_a}\dim_{\theta_n}}}{\dim_\iota}\cdot
  U_n(a)_{ji}$ are unitary. Explicitly, 
\bea\label{inv}
\sumno_a\dim_{\gamma_a}\dim_{\theta_n} \ol{U_n(a)_{ji}}
U_{n'}(a)_{j'i'} = \delta_{nn'}\delta_{ii'}\delta_{jj'}\cdot
(\dim_\iota)^2. \hskip-10mm
  \eea 
\end{lemma}
\eref{inv} may be read as the orthogonality of the functions 
$a\mapsto U_n(a)_{ji}$ with respect to the inner product 
induced by the Haar measure 
$$(f,g)_H= \big(\sumno_a \mu_a\big)\inv\sumno_a \ol{f(a)}\mu_ag(a).$$
The corresponding Hilbert space carries the regular representation 
$$(R_bf)(a) = \sumno_c \lambda_{ab}^cf(c).$$

\medskip

{\em Proof of \pref{p:pw}:} Let $k_a\in K_{N\subset M}$ stand for the
abstract basis elements of the hypergroup defined by the composition
law \eref{coeff} of the positive maps $\phi_a$. 
The linear map defined on the basis by $k_a\mapsto
\bigoplus_nU_n(a)$ is an algebra homomorphism $K_{N\subset M}\to
\bigoplus_n\Mat_{m_n}(\CC)$. It is bijective, hence an isomorphism by
\lref{l:uni}. Under this isomorphism, the representations $U_n$ pass to
the restrictions of $\bigoplus_n\Mat_{m_n}(\CC)$ to $\Mat_{m_n}(\CC)$.
Since the latter are the inequivalent irreducible matrix
representations of $\bigoplus_n\Mat_{m_n}(\CC)$, the claim follows.  
\qed

\section{Boundary conditions}
\label{s:bcond}
\setcounter{equation}{0}

We now turn to the application of \sref{s:hyperg} to boundary
conditions in quantum field theory, as exposed in \sref{s:i-qft}.

\subsection{General setup}
\label{s:setup}

We consider a boundary separating two local quantum field theories
$B^L$ and $B^R$ with a common subtheory $A$. Locality ensures that the
Q-systems $Q^L$ for $A(O)\subset B^L(O)$ and $Q^R$ for $A(O)\subset
B^R(O)$ are commutative \cite{LR}, with the braiding of the dual
canonical endomorphisms given by the DHR braiding of $A$.  

To comply with the physical interpretation as a transparent boundary,
the relative algebraic position of the local algebras $B^L(O)$ and
$B^R(O)$ must fulfil several algebraic conditions. A ``boundary
condition'' is a realization of these conditions on a Hilbert
space. The conditions can be cast into the following form
\cite{BKLR}: For each $O$ one has a subfactor $A(O)\subset C(O)$ with
intermediate embeddings $\jmath^L$ of $B^L(O)$ and $\jmath^R$ of
$B^R(O)$: 
\bea
\label{diagram}
\begin{array}{ccccc}
&&B^L(O)&&\\[-4mm]
&\stackrel{\iota^L}\uphook&&\stackrel{\jmath^L}\downhook&
\\[-2mm] A(O) &&&& C(O)\\[-2.4mm]
&\stackrel{\iota^R}\downhook&&\stackrel{\jmath^R}\uphook&
\\[-2.4mm] &&B^R(O)&&
\end{array}
\eea
Thus, one has\footnote{The embeddings make
  \eref{diagram} a commuting diagram, but not a commuting square in
  the sense of \cite{GHJ,EK}, because in general
  $\jmath^L(B^L(O))\cap\jmath^R(B^R(O))\neq A(O)$.} 
\bea
\label{interm}
\jmath^L\circ\iota^L = \jmath^R\circ\iota^R.
\eea
In addition, the crucial requirement is locality: $\jmath^L(B^L(O_1))$
must commute with $\jmath^R(B^R(O_2))$ whenever $O_1\subset O$ lies to the
spacelike left of $O_2\subset O$ (``left-locality''). 

Finally, one has $C(O)=\jmath^L(B^L(O))\vee\jmath^R(B^R(O))$. (If this
is not the case, we call $C$ a defect; a general theory of defects is
still under investigation \cite{BR}.) The boundary condition is called
irreducible, if the subfactor $A(O)\subset C(O)$ is irreducible. 

The condition \eref{interm} is nontrivial even if $B^L(O)=B^R(O)\equiv B(O)$
and $\iota^L=\iota^R$, because still $\jmath^L$ may be different from
$\jmath^R$. In this case, we have two different embeddings of the same
algebra into $C(O)$; the boundary condition then specifies the way in
which $\jmath^L$ and $\jmath^R$ may differ. It is our main result in
\sref{s:homo} that they differ (in a way to be detailed) by a positive map
$\phi_a$ representing an element $k_a$ of the hypergroup $K$
associated with $A\subset B$, as mentioned above. In particular, if
$A(O)=B(O)^G$ are the fixed points under a global gauge group, one has
$K=G$, and the two embeddings differ by a gauge transformation. 

We shall establish in \sref{s:jux} that the composition law \eref{hg}
of the hypergroup describes the juxtaposition of two boundaries,
making contact with \cite{R}.   

In \cite{BKLR}, we have defined a ``universal construction'' as a
solution to \eref{diagram}, where $C(O)=C\uni(O)$ is not a
factor. (For the necessary adaptation to the case $N\subset M$ 
where $M$ is not a factor, cf.\ \cite[Sect.~2.3]{Tcats}.) 
The central decomposition of $C\uni(O)$ gives rise to irreducible
solutions to \eref{diagram} (= boundary conditions), and every
boundary condition arises in this decomposition. 

One should, of course, construct $A(O)\subset C\uni(O)$ for every
$O$ (interval or doublecone). Here, we just sketch the construction 
for one fixed region $O$. One has $C\uni(O)=\iota\uni(A(O))\cdot V\uni$, 
where the dual canonical endomorphism $\Theta\uni=\ol\iota\uni\iota\uni$ 
is the restriction to $A(O)$ of a DHR endomorphism of the net $A$
localized in $O$. The extension to arbitrary regions $O_1$ proceeds 
by a standard method: $C\uni(O_1)=\iota\uni(A(O_1))\cdot
\iota\uni(u)V$ where $u\in A$ is a unitary charge transporter such
that $\Ad_u\circ\Theta\uni$ is localized in $O_1$. 

The universal construction proceeds by the specification of the Q-system 
for $A(O)\subset C\uni(O)$ as a braided product of the given Q-systems
$Q^L$ and $Q^R$. It is defined by $\Theta\uni = \theta^L\theta^R$, and
\bea\label{bprod}
W\uni=w^L\times w^R\quad\hbox{and}\quad X\uni=(1_{\theta^L}\times
\eps^-_{\theta^L,\theta^R}\times 1_{\theta^R})\circ (x^L\times x^R).\hskip-15mm
\eea
Here, the requirement of left-locality determines the correct choice
of the braiding $\eps^-_{\rho,\sig}=\eps_{\sig,\rho}^*$ in the 
conventions of \cite{FRS}, such that $\eps^-_{\rho,\sig}=1$ whenever $\rho$
is localized to the left of $\sig$. We write $Q\uni=Q^L\times^-Q^R$. 

The commutativity of $Q^L$ and $Q^R$ implies \cite[Prop.~4.19]{BKLR} that 
$$C\uni(O)'\cap C\uni(O) = A(O)'\cap C\uni(O).$$
This ensures that the central decomposition of $C\uni(O)$ gives rise to
irreducible boundary conditions, each of them arising in a
representation with a unique vacuum vector.

Let $V\uni\in C\uni(O)$ be the canonical generator of the extension
$A(O)\subset C\uni(O)$, such that $(W\uni,V\uni)$ is a solution to
the conjugacy relations. The canonical generators of the intermediate
embeddings of $B^L(O)$ and $B^R(O)$ are respectively 
\bea\label{emb}
\jmath^L(v^L)=\iota\uni(1_{\theta^L}\times w^R{}^*)\!\cdot\!
V\uni, \quad \jmath^R(v^R)=\iota\uni(w^L{}^*\times
1_{\theta^R})\!\cdot\! V\uni.\hskip-15mm
\eea

We have in \cite{BKLR} characterized the minimal central projections
of $C\uni(O)$ as $E_a=\iota\uni(q_a)V\uni$, where 
$q_a\in \Hom(\Theta\uni,\id_{C\uni})$ satisfy $(q\times q)\circ X\uni=q$ 
and $q^*=(q\times 1_{\Theta\uni})\circ X\uni W\uni = (1_{\Theta\uni}\times q)\circ
X\uni W\uni$. By Frobenius reciprocity, $q_a\in\Hom(\theta^L\theta^R,\id)$ 
are in bijection with intertwiners 
$$I_a=\varphi(q_a):=(1_{\theta^L}\times q_a)\circ (x^Lw^L\times 1_{\theta^R})
\in\Hom(\theta^R,\theta^L),$$
and $I_a$ are minimal projections of $[\Hom(\theta^R,\theta^L),\ast]$
equipped with the generalized $\ast$-product (cf.\ \sref{s:hetero}) 
\bea\label{astRL} x^L{}^*\circ (t_1\times t_2)\circ x^R.
\eea

\subsection{The hypergroup of boundary conditions}
\label{s:homo}

Let us first concentrate on the case $Q^L=Q^R$, i.e., the local
extensions $A\subset B^L$ and $A\subset B^R$ are isomorphic. The
boundary conditions under consideration are therefore boundary
conditions between two isomorphic copies of a quantum field theory
$B$. One may think of
them as ``discontinuities'' of $B$, not affecting the subtheory
$A$. In a Euclidean setting, one would think of a transition line
between two different phases of continuum statistical system, the
simplest nontrivial example being the ordered and the disordered phase
of the critical Ising model \cite{ST,BKLR}. 

In this case, we shall suppress the superscripts $L,R$ for the
extensions $\iota:A\subset B$; but it is crucial to retain the
superscripts for the intermediate embeddings $\jmath^L\neq \jmath^R$
of $B$ into $C$. We thus have 
$\iota\uni=\jmath^L\circ\iota=\jmath^R\circ\iota$. 

Turning to the characterization of irreducible boundary conditions as above, we
have to find the minimal projections $I_a\in [\Hom(\theta,\theta),\ast]$.  

The following obvious conclusion was not explicitly mentioned in
\cite{Tcats,BKLR}: The minimal projections $I_a\in \Hom(\theta,\theta)$ 
with respect to the $\ast$-product are mapped by the Fourier transform
\eref{fou} to the minimal projections $e_a$ of $\Hom(\gamma,\gamma)$
with respect to the ordinary = concatenation product. The latter are
the projections $e_a=s_as_a^*$ as in \sref{s:hyperg} onto the
irreducible subsectors $\gamma_a\prec\gamma$ that are all contained
with multiplicity 1 by \pref{p:mfree}.  
\begin{lemma} \label{l:eIE} The minimal central projections of the universal
construction are obtained by the chain of maps
$$e_a\stackrel{\chi\inv}\longmapsto I_a \stackrel{\varphi\inv}\longmapsto q_a
\mapsto E_a = \iota\uni(q_a)V\uni.$$
Equivalently, $\ol\iota\uni(E_a)=\ol\iota(e_a)$ as elements of $A$.  
\end{lemma}
{\em Proof:} We compute
$$\ol\iota\uni(E_a)=\theta^2(q_a)X\uni =
\theta(q_a)x\theta(x)=x^*\theta(I_a)\theta(x) = \ol\iota(e_a).$$
The first equality follows from $\Theta\uni=\theta^2$ and
$\ol\iota\uni(V\uni)=X\uni$. The second uses the definition
\eref{bprod} of $X\uni$ and the commutativity of the Q-system (i.e.,
locality of the QFT $B$). The third uses $q_a=\varphi\inv(I_a)$ and
the relations of the Q-system. The last equality uses $I_a=\chi\inv(e_a)$
and $x=\ol\iota(v)$. \qed 

\medskip

The following proposition relates the central projections $E_a$ to the
action of the 
hypergroup $K_{A\subset B}$ associated with the subfactor $A(O)\subset
B(O)$ on the isometries $\psi_{n,i}\in\Hom(\iota,\iota\theta_n)$ as in
\sref{s:reps}. 
\begin{propo}\label{p:proj} The following equivalent expansions hold: 
\bea\label{main}
\jmath^L(\psi_{n,j}^*)\jmath^R(\psi_{n,i}) = \sumno_a U_n(a)_{ji}\cdot E_a.
\eea
\bea\label{Eahg} E_a=
\sumno_{n,i}\frac{\dim_{\gamma_a}\dim_{\theta_n}}{(\dim_\iota)^2}\cdot
\jmath^L\phi_a(\psi_{n,i}^*)\jmath^R(\psi_{n,i}).
\eea
\end{propo}
Before we sketch the proof of these formulae, let us explain how they
specify the irreducible boundary conditions selected by the central
projections $E_a$: In the quantum field theory $B$, 
$\psi_{n,i}$ are ``charged field operators'' \cite{BKLR} that
intertwine the vacuum representation of $A$ with the DHR
representations $\theta_n$ of $A$ within the vacuum representation of
$B$. They are embedded via \eref{emb} as $\jmath^L(\psi_{n,i})$ and
$\jmath^R(\psi_{n,i})$ into the universal construction $C\uni$, and
the operators $\jmath^L(\psi_{n,j}^*)\jmath^R(\psi_{n,i})$ span the center of
$C\uni$ \cite{BKLR}. Thus, they take numerical values in 
irreducible representations. 

The irreducible boundary condition $a$ is the representation
of $C\uni$ in which $\pi_a(E_b)=\delta_{ab}\cdot 1$. Thus one obtains the
numerical values 
\begin{coro}\label{c:inv}
\bea\label{solve}
\pi_a(\jmath^L(\psi_{n,j}))^*\cdot\pi_a(\jmath^R(\psi_{n,i}))
= U_n(a)_{ji}.
\eea
\end{coro}
In special cases, when this matrix of inner products happens to be
unitary, the Cauchy-Schwarz inequality implies linear relations 
between the embedded field operators $\pi_a(\jmath^L(\psi_{n,i}))$ and
$\pi_a(\jmath^R(\psi_{n,i}))$. In particular, the trivial subsector
$\gamma_0=\id_B$ $\RA$ $\phi_0=\id_B$ $\RA$
$\pi_a(\jmath^L(\psi_{n,j}))^*\cdot\pi_a(\jmath^R(\psi_{n,i}))=\delta_{ij}$
implies the trivial boundary condition
$\jmath^L(\psi_{n,i})=\jmath^R(\psi_{n,i})$. 
In the general case, a boundary condition is a sesquilinear relation,
specifying ``angles'' between the two spaces of embedded charged field
operators. In particular, a linear relation of the form 
``$\pi\jmath^R(\psi) = \pi\jmath^L(\phi(\psi))$'' is in general not
true, and in extreme cases, $\jmath^R(\psi)$ and $\jmath^L(\psi)$ are
linearly independent in a given representation. 

We indicate how the linearization works out in the case when $A$ is a
fixed-point subnet of $B$ under some finite global gauge group $G$. In
this case, the subsectors of $\gamma$ are the gauge automorphisms
$\gamma_g$ of $B$, and $\phi_g=\gamma_g$ and the hypergroup is
$K_{A\subset B}=G$. Therefore, $U_n(g)_{ji}$ are in fact the unitary
matrix representations of $G$, and \lref{l:uni} becomes a familiar
identity (Peter-Weyl). The scalar products
$$\pi_g\big(\jmath^L(\psi_{n,j}^*)\jmath^R(\psi_{n,i})\big) = U_n(g)_{ji}$$ 
imply $X^*X=0$ for $X=\pi_g\jmath^R(\psi_{n,i})-\sumno_j\pi_g\jmath^L(\psi_{n,j})U_n(g)_{ji}$. Hence 
$$\pi_g\jmath^R(\psi_{n,i})= \sumno_j\pi_g\jmath^L(\psi_{n,j})U_n(g)_{ji} =
\pi_g\jmath^L(\gamma_g(\psi_{n,i})).$$
In other words: the boundary condition states that the left and right
embeddings of $B$ differ by the gauge transformation $\gamma_g$. 

\medskip

{\em Proof of \pref{p:proj}:} \eref{main} and \eref{Eahg} are clearly
equivalent by \eref{rep} and \lref{l:uni}. We shall prove \eref{main}.

It is convenient to prove the equality after application of
$\ol\iota\uni$. On the left-hand side, we use \eref{emb} to compute
the embeddings of 
the left and right charged fields, as well as the definition of
$X\uni=\ol\iota\uni(V\uni)$. We find 
$$\ol\iota\uni\big(\jmath^L(\psi_{n,j}^*)\jmath^R(\psi_{n,i})\big) =
x^*\theta(\eps_{\theta,\theta}\theta(w_{n,j}w_{n,i}^*)x).$$
By commutativity of the Q-system, this equals
$$\ol\iota\uni\big(\jmath^L(\psi_{n,j}^*)\jmath^R(\psi_{n,i})\big)
=x^*\theta(w_{n,j}w_{n,i}^*)\theta(x) =
\ol\iota\big(\chi(w_{n,j}w_{n,i}^*)\big).$$ 
On the right-hand side, we use \cref{l:eIE} and \eref{U}, written as 
$s_a^*\circ\chi(w_{n,j}w_{n,i}^*)\circ s_a$. The sum
over $a$ can be trivially performed: 
$$\sumno_a U_n(a)_{ji}\cdot\ol\iota\uni(E_a) = \ol\iota\Big(\sumno_a s_as_a^* \chi(w_{n,j}w_{n,i}^*) s_as_a^*\Big)=\ol\iota\big(\chi(w_{n,j}w_{n,i}^*)\big).$$
This concludes the proof. \qed

\subsection{Composition of boundary conditions}
\label{s:jux}

In \cite{R}, one of us has defined a composition of boundary conditions which
may be understood in physical terms as the boundary conditions between
a QFT to the left of strip $a<x<b$ and a QFT to the right of the strip
that are, as sesquilinear relations or as angles between spaces of
charged field operators, compatible with a pair of boundary conditions
with a third QFT inside the strip (``juxtaposition of boundaries''). 
This composition is defined by the concatenation product of
intertwiners $I_a\in\Hom(\theta,\theta)$ as in \cref{l:eIE}: 
$$I_aI_b = \sumno_{c}C_{ab}^c\, I_c.$$
Under the Fourier transform \eref{fou} this amounts to the
convolution product of the projections $e_a\in\Hom(\gamma,\gamma)$:  
$$e_a\ast e_b = \sumno_{c}C_{ab}^c\, e_c,$$
and because the convolution product of two positive operators is
positive, it is clear that $C_{ab}^c$ are non-negative numbers. 
\begin{propo}\label{jux}
Up to a trivial rescaling, this composition of boundary conditions is
isomorphic to the hypergroup product of $K_{A\subset B}$. 
\end{propo}
{\em Proof:} It is sufficient to go back to the proof of (ii) of
\pref{p:hg}. Let $s_{ab}\circ t_c = \alpha_{ab}^c\cdot
s_c$. Then, by inserting the decomposition of unity
$1_{\gamma_a\gamma_b}=\sum_i t_it_i^*$ into $e_a\ast e_b$, one finds
$e_a\ast e_b=\sum_c \vert \alpha_{ab}^c\vert^2 e_c$, i.e.,
$\dim_\iota\cdot e_a$ satisfy the same composition law \eref{coeff} as $\phi_a$. 
\qed 

\medskip

{\em Examples:} The GHJ subfactor (cf.\ \sref{s:subf}) describes the
$E_6$ conformal embedding $A(O)\subset B(O)$ of the chiral QFTs
$SU(2)_{10}\subset SO(5)_1$ \cite{RST}. The nontrivial charged field
operator $\psi$ carries the DHR charge $s=3$ of $\SU(2)_{10}$. The model
admits one nontrivial boundary condition, for which the angle between
$\jmath^L(\psi)$ and $\jmath^R(\psi)$ can be computed from the
nontrivial one-dimensional representation of the hypergroup $K_{\rm
  GHJ}$, $U(1)=-1/d_1=\sqrt3-2$. \\ 
The following examples have been worked out before we recognized that
the common underlying structure is a hypergroup:\\  
In \cite{BKLR}, we presented the case of the two-dimensional
relativistic Ising model as a prototype. Apart from the trivial
boundary condition and the ``fermionic'' boundary condition that is a
$\ZZ_2$ gauge transformation, there is a third ``dual'' boundary
condition whose composition is a convex sum (mixture) of the trivial
and fermionic ones. The dual boundary condition is the relativistic
analogue of the phase boundary in which the ordered and the disordered
phase of the critical Euclidean Ising model coexist along a line \cite{ST}. \\  
Czechowski \cite{Cz} has computed the intertwiners $I_a$ by
diagonalizing the convolution product for Q-systems in several DHR
categories that contain $\Rep(S_3)$, and has determined their composition. \\
For the canonical \cite{LR,Tcats} two-dimensional extension of the
tensor product of left- and right-moving chiral QFTs $A\otimes A$
along a subcategory $\Rep(G)\subset \DHR(A)$, one finds $K_{A\otimes
  A\subset B} = \Conj(G)$ \cite{R}. \\ For the full center extensions
$A\otimes A\subset B$ associated with a chiral Q-system $q$ in
$\DHR(A)$, where the chiral theory $A$ has modular
tensor category $\DHR(A)$, one finds that $K_{A\otimes A\subset B}$ is
isomorphic to the hypergroup of the fusion category of $q$-$q$
bimodules \cite{Tcats,BKLR}. This was first discovered, in a different
setup, in \cite{FFRS}.

\subsection{The hypergroupoid of boundary conditions}
\label{s:hetero}

The picture developped in the previous section naturally and with
almost no effort
generalizes to the case when the observables on both sides of the
boundary are different extensions of $A(O)$. In this case, one arrives
at a hypergroupoid whose objects $X$ correspond to inequivalent
extensions $A(O)\subset B^X(O)$, and whose arrows
$k^{XY}_a\in\Mor(Y,X)$ satisfy the obvious generalization of
\eref{hg}. They are represented by completely positive maps  
$\phi^{XY}_a:B^Y(O)\to B^X(O)$. 

For a boundary between QFTs $B^L$ to the left and $B^R$ to the right,
we have $X=L$, $Y=R$. The Fourier transform \eref{fou}
generalizes to a map $\chi:\Hom(\theta^R,\theta^L)\to
\Hom(\iota^L\ol\iota^R,\iota^L\ol\iota^R)$. Correspondingly, the
convolution product on $\Hom(\theta^R,\theta^L)$ is given as in 
\eref{astRL}. 

The universal construction is given by the braided product of
Q-systems, as in \sref{s:setup}. Its minimal central projections $E_a$
are obtained as in \cref{l:eIE}, with the obvious replacements
$\gamma_a\prec \iota^L\ol\iota^R$, $e_a\in
\Hom(\iota^L\ol\iota^R,\iota^L\ol\iota^R)$, $I_a\in
\Hom(\theta^R,\theta^L)$, and $q_a\in
\Hom(\theta^L\theta^R,\id)$. Finally, \pref{p:proj} and \cref{c:inv}
become
$$E_a=
\sumno_{n,i}\frac{\dim_{\gamma_a}\dim_{\theta_n}}{\dim_{\iota^L}\dim_{\iota^R}}\cdot
\jmath^L\phi_a(\psi^L_{n,i}{}^*)\jmath^R(\psi^R_{n,i}),
$$
where $n$ runs over all joint irreducible subsectors $\theta_n$ of $\theta^L$ and
$\theta^R$, and
$$
\pi_a(\jmath^L(\psi^L_{n,j}))^*\cdot\pi_a(\jmath^R(\psi^R_{n,i}))
= U^{LR}_n(a)_{ji}
$$ 
with 
$$U^{LR}_n(a)_{ji}\cdot 1_{\gamma_a} = (v^L{}^*\times s_a^*)\circ(1_{\iota^L}\times
w^L_{n,j}w^R_{n,i}{}^*\times 1_{\ol\iota^R})\circ (s_a\times v^R). $$

Thus, the boundary conditions are sesquilinear relations
between the charged field operators of $B^L$ and $B^R$ carrying common charges. 

\appendix

\section{A result on the quantization of dimensions}
\label{app}
\setcounter{equation}{0}

Let us return to the central decomposition of the braided product 
$Q\times^-Q$ of a commutative Q-system for $A\subset B$. It is defined
by the minimal projections $E_a\in C\uni'\cap C\uni$ induced by the
minimal projections $e_a\in\Hom(\gamma,\gamma)$ whose range is
$\gamma_a\prec\gamma$, as in \sref{s:homo}. The result is, for each
$a$, an intermediate embedding
$$A\stackrel\iota \hookrightarrow B\stackrel{\jmath_a}\hookrightarrow C_a=E_aC\uni$$
whose dual canonical endomorphism
$\theta_a=\ol\iota\ol\jmath_a\jmath_a\iota$ equals
$\ol\iota\gamma_a\iota\prec \ol\iota\gamma\iota=\theta^2$. This
equality does not imply that $\gamma_a=\ol\jmath_a\jmath_a$ (because
$\iota$ is not surjective), but it implies $\dim_{\gamma_a}=
\dim_{\ol\jmath_a\jmath_a} = (\dim_{\jmath_a})^2=[C_a:B]$. 
Since the net of subfactors is entirely irrelevant in this argument,
we have proven: 
\begin{propo}\label{p:app} 
If $N\subset M$ is an irreducible finite-index subfactor
  with a commutative Q-system, then the dimension $\dim(\gamma_a)$ of
  every irreducible sub-endomorphism $\gamma_a$ of the canonical
  endomorphism $\gamma\in\End_0(M)$ equals the index of some subfactor. 
\end{propo}
The statement in the conclusion is nontrivial, because in general,
a dimension is a square root of an index, and the index is quantized
below 4. The statement is trivially true for a fixed point
subfactor $N=M^G$ under an outer action of a finite group, because all
$\gamma_a$ are automorphisms; it is also
obvious for the canonical subfactor of a modular braided tensor category
(also known as Longo-Rehren = Jones-Wassermann = two-interval = Cardy
subfactor), because these subfactors are anti-self-dual, hence the
sector decomposition of $\gamma$ is isomorphic to that of
$\theta\cong\bigoplus_\rho \rho\otimes\ol\rho$ with dimensions
$\dim_{\rho\otimes\ol\rho}=(\dim_\rho)^2=[N:\rho(N)]$. It can also be
easily verified, e.g., for the GHJ subfactor (which is commutative
because it describes the $E_6$ conformal embedding in conformal QFT)
where the nontrivial subsector has $\dim(\gamma_a)=2+\sqrt3 =
(2\cos\frac\pi{12})^2$. To our knowledge, the statement of \pref{p:app} was not known to be a
general fact.  
\begin{coro} The index of an irreducible subfactor $N\subset M$ with
commutative Q-system equals 1 plus a sum of indices $[L_a:M]$. 
\end{coro}

This observation gives an obvious a priori quantization of the index
below 5, with only one non-integer value below 4, namely
$(2\cos(\frac\pi{10}))^2=1+(2\cos(\frac\pi5))^2 =
\frac12(5+\sqrt5)\approx 3.618$, in accord
with \cite[Thm.~2.3]{CKL}.


\end{document}